\documentclass[showpacs,preprintnumbers,amsmath,amssymb,preprint,aps,pre]{revtex4}

\usepackage{graphics,graphicx}

\usepackage{epsfig}
\begin{document}

\title{Role of delay in the stochastic creation process}
\author{L. F. Lafuerza, R. Toral}

\affiliation{IFISC, Instituto de F{\'\i}sica Interdisciplinar y Sistemas Complejos, CSIC-UIB,  Campus UIB, E-07122 Palma de Mallorca, Spain}

\date{\today}


\begin{abstract}
We develop an approximate theoretical method to study discrete stochastic birth and death models that include a delay time. We analyze the effect of the delay in the fluctuations of the system and obtain that it can qualitatively alter them. We also study the effect of distributed delay. We apply the method to a protein-dynamics model that explicitly includes transcription and translation delays. The theoretical model allows us to understand in a general way the interplay between stochasticity and delay.
\end{abstract}

\maketitle
\section{Introduction}

Fluctuations play an important role in many areas of science\cite{VK}, and their study has become a well defined discipline. Delay in the interactions is also a common phenomenon in natural and artificial systems, and it is well known that it can alter qualitatively the dynamical behavior, for example inducing oscillations or even chaos\cite{makeyglass}. In particular, both fluctuations and delay are relevant in gene-regulation systems, where a lot of research effort, both theoretical and experimental, has been recently carried out\cite{elowitz,oudenaarden,lewis,Swinburne,paulsson}. 

The connived effect of stochasticity and delay is not completely understood. In this context, many studies have focused in delayed stochastic, Langevin, differential equations or Fokker-Planck equations \cite{langevindelay,Frank} that assume continuous variables, or random walks in discrete time\cite{rwdelay}. Stochastic models with continuous time but discrete variables are the natural description of many systems such as chemical reactions, population dynamics, epidemics, etc. In some cases this discreteness is a major source of fluctuations \cite{aparicio}.

In this work, we study some general stochastic birth and death processes that include delay. We follow a master equation approach that considers discrete variables in continuous time. We present an analytical treatment that allows us to study the effect of delay  and show that the delay can alter qualitatively the character of the fluctuations. We also consider the situation with distributed delay and study how the fluctuations change as the delay distribution becomes wider. 

The paper is organized as follows: In the following section \ref{sec:2} we present the theoretical approach, applying it to a general one-step birth-death model and discuss the influence of the delay in the fluctuations of this system. In section \ref{sec:3} we consider the effect of distributed delay. In section \ref{sec:4} we study a two-step transcription-translation model, more relevant to gene regulation. We finish in section \ref{sec:5} with some conclusions and comments.

\section{Stochastic creation with delay}
\label{sec:2}

Let us start by considering a simple one-step stochastic process in which the number of units (e.g. molecules) $n$ of some compound $X$ can only increase or decrease by one:
\begin{eqnarray}
\emptyset {{C \atop \longrightarrow}\atop{}} X, \hspace{0.5cm} X {{D \atop \longrightarrow}\atop{}} \emptyset.
\end{eqnarray}
The creation, $C(n)$, and annihilation, $D(n)$, rates depend, in general, on $n$.  The probability $P(n,t)$ that there are $n$ molecules at time $t$ follows a master equation\cite{VK}:
\begin{equation}
\label{master_nodelay}
 \frac{\partial P(n,t)}{\partial t}=(E-1)[D(n)P(n,t)]+(E^{-1}-1)[C(n)P(n,t)],\,t\ge 0
\end{equation}
being $E$ the step operator: $E^k[f(n)]=f(n+k)$.

Our main aim in this paper is to consider that the creation of an $X$ particle takes a finite amount of time $\tau$. More specifically, we consider that the creation is a stochastic process initiated at a rate $C(n)$ but, once initiated, it takes a finite amount of time $\tau$ to be completed. Schematically:
\begin{eqnarray}
\label{eq:process}
\emptyset {{C \atop \longrightarrow}\atop{}} X^*, \hspace{0.5cm} X^*{{ {}\atop \Longrightarrow}\atop{\tau}}X,\hspace{0.5cm} X {{D \atop \longrightarrow}\atop{}} \emptyset.
\end{eqnarray}
Here, $X^*$ is considered to signal the beginning of the process that, after a time $\tau$, will lead to $X$. The creation of  $X^*$ is a stochastic process, but the step leading from $X^*$ to $X$ is deterministic, requiring a constant time $\tau$ for completion (this is indicated by a double arrow, while a single arrow denotes an stochastic event). We consider that the stochastic variable $n$ takes into account just the number of $X$ molecules (not including $X^*$). Similar, although not identical, processes have been considered before in the context of protein synthesis\cite{BVTH-2005, Galla-2009}. In a very simple manner, we can think that $X^*$ indicates the beginning of the transcription process of a protein from a gene,  but that once initiated the transcription plus translation steps take a time $\tau$ to be completed. In this case, the creation rate $C(n)$ depends on $n$ if there is auto inhibition or auto activation, leading to a negative or positive, respectively, feedback loop. 

Due to the presence of delay, the master equation of the process involves now the two-times probability distribution $P(n,t;n',t')$ as:
\begin{eqnarray}
 \frac{\partial P(n,t)}{\partial t}&=&(E-1)[D(n)P(n,t)]+(E^{-1}-1)\left[\sum_{n'=0}^{\infty}C(n')P(n',t-\tau;n,t)\right],\label{master_delay}
\end{eqnarray}
valid for $t\ge0$. The creation term takes into consideration that the probability that a particle is created at time $t$ is the sum of all contributions in which there were $n'$ particles at time $t-\tau$ and an $X^*$ particle was created at that time, at a rate $C(n')$, leading necessarily at time $t$ to an $X$ particle. This equation is the basis of our subsequent analysis. We will eventually consider, for the sake of concreteness, that the annihilation of particles occurs through independent events at individual rate $\gamma$ and, hence, $D(n)=\gamma n$. Formally, equation (\ref{master_delay}) can be written in the form:
\begin{equation}
\label{master_effdelay}
 \frac{\partial P(n,t)}{\partial t}=(E-1)[D(n)P(n,t)]+(E^{-1}-1)[\tilde C(n,t)P(n,t)],
\end{equation}
with an effective time dependent rate
\begin{equation}\label{eq:ctilde}
\tilde C(n,t)=\langle C(n'),{t-\tau}|n,t\rangle,
\end{equation}
where we have introduced the notation for the conditional average $\langle f(n_1),t_1|n_2,t_2\rangle=\sum_{n_1}f(n_1)P(n_1,t_1|n_2,t_2)$. Other, non-conditional averages, will be denoted as  $\langle f(n)\rangle_t=\sum_n f(n)P(n,t)$. The conditional average $\langle n,t|k,t_0\rangle$ satisfies the  evolution equation, valid for $t\ge 0$:
\begin{eqnarray}
\frac{d\langle n,t|k,t_0\rangle}{dt}&=&-\gamma\langle n,t|k,t_0\rangle+\langle C(n),t-\tau|k,t_0\rangle,\label{averagedelaygen}
\end{eqnarray}
with initial condition $\langle n,t_0|k,t_0\rangle= k$. Higher order averages obey a hierarchy of equations which we do not  need to write down for the purposes of this paper. The resolution of this hierarchy would allow one to compute the average value $\langle f(n)\rangle_{t}$ of any function $f(n)$ which can be expanded as a Taylor series of $n$.

We will be mostly interested in the steady-state, where the averages $ \langle f(n)\rangle_{st}\equiv \lim_{t\to\infty}\langle f(n)\rangle_t$ are time independent and the conditional averages depend only on the time difference, $\langle f(n_1),t|n_2\rangle_{st}\equiv\lim_{t'\to\infty}\langle f(n_1),t+t'|n_2,t'\rangle$. They can be computed, respectively, from the steady-state probability distributions $P_{st}(n)=\lim_{t\to\infty}P(n,t)$ and $P_{st}(n_1,t|n_2)=\lim_{t'\to\infty}P_{st}(n_1,t'+t|n_2,t')$. Formally, the knowledge of the steady value $\tilde C_{st}(n)\equiv \lim_{t\to\infty}\langle C(n'),t-\tau|n,t)\rangle=\langle C(n'),-\tau|n\rangle_{st}$, allows the calculation of the steady-state probabilities $P_{st}(n)$, after imposing $\frac{\partial P(n,t)}{\partial t}=0$ in Eq.(\ref{master_effdelay}),  as \cite{VK}:
\begin{equation}\label{probgeneral}
 P_{st}(n)=P_{st}(0)\prod_{k=0}^{n-1}\frac{\tilde C_{st}(k)}{D(k+1)}=\frac{P_{st}(0)}{\gamma^nn!}\prod_{k=0}^{n-1}\tilde C_{st}(k),
\end{equation}
$P_{st}(0)$ is fixed by the normalization condition. In the following subsections, we will discuss two methods to obtain the conditional averages needed for the calculation of $\tilde C_{st}(n)$. This will allow us to obtain the steady state probabilities as well as the mean value $\langle n\rangle_{st}$, variance $\sigma^2_{st}$ and correlations $K(t)=\langle n\langle n',t|n\rangle_{st}\rangle_{st}-\langle n\rangle_{st}^2$. 

\subsection{The independent-times approximation}

The first method assumes that the conditional average values do not depend on previous history or, equivalently, that the two-times probability distribution factorizes as $P(n_1,t_1;n_2,t_2)=P(n_1,t_1)P(n_2,t_2)$. This implies that in Eq.({\ref{eq:ctilde}) we can set $\tilde C(n,t)=\langle C(n')\rangle_{t-\tau}$, independent on $n$. On empirical grounds, it is expected that this approximation will be valid for large $\tau$ where the events at $t$ and $t-\tau$ can be considered independent, although we will show later that this is not the case. In the steady state, this assumption implies $\tilde C(n,t)=\langle C(n)\rangle_{st}$, a constant. Replacing this result in Eq.(\ref{probgeneral}) we obtain that  the steady-state follows a Poisson distribution $P_{st}(n)=e^{-\chi}\frac{\chi^n}{n!}$ with $\chi=\frac{\langle C(n)\rangle_{st}}{\gamma}$. The, yet unknown, steady state average value is obtained through the consistency relation $\langle C(n)\rangle_{st} =\sum_{n} C(n)P_{st}(n)$. Once this equation is solved, the mean and variance of the distribution follow: $\langle n\rangle_{st}=\sigma^2_{st}=\frac{\langle C(n)\rangle_{st}}{\gamma}$.

The consistency relation can be explicitly solved in the linear case $C(n)=c-\epsilon n$ with the result $\langle n\rangle_{st}=\frac{c}{\epsilon+\gamma}$. In the literature, and in the field of protein transcription, it is usually considered a negative feedback loop where the creation rate is a decreasing, non linear function, for example $C(n)=\frac{c}{1+\epsilon n}$. This corresponds to a gene repressed directly by the protein it encodes for, in the limit where the binding and unbinding of this protein is fast compared to the rest of time scales of the system. The approximation is good if the unbinding rate of the protein from the promotor is much greater (or the order of ten times) than the degradation rate of the protein\cite{hornos}. In this case, the consistency relation reduces to:
\begin{equation}
\langle n\rangle_{st}=\frac{c}{\gamma}\int_0^{\infty} dx\,\exp[-x+\langle n\rangle_{st}(e^{-\epsilon x}-1)],
\end{equation}
which, in general, needs to be solved numerically. In the limit $\epsilon\to 0$ we can expand $e^{-\epsilon x}-1=-\epsilon x$ to derive $\langle n\rangle_{st}=\frac{c/\gamma}{1+\epsilon \langle n\rangle_{st}}$, the mean-field result.
Since at the steady state the effective creation rate $\tilde C(k,t)=\langle C(n)\rangle_{st} $ is constant, the process is a simple birth-death process and the correlations decay exponentially as $K(t)=\sigma^2_{st}e^{-\gamma |t|}$.

Within this independent-times approximation, the steady state average value $\langle n\rangle_{st}$ and variance $\sigma^2_{st}$ are equal (Poisson distribution) and do not depend on the delay time $\tau$. This is a general result that does not depend on the specific functional form for the creation rate $C(n)$. As discussed before, this is naively expected to hold in the case of a large delay $\tau$. In the numerical simulations, however, it is observed that the fluctuations are sub-Poissonian for small $\tau$ and super-Poissonian for large $\tau$. The details of the simulations for this stochastic process including delay are given in the Appendix 1. Note that the case $\tau=0$ can be solved (exactly) by a variety of methods. Within our treatment, and according to Eq.(\ref{eq:ctilde}), for $\tau=0$ the conditional probability is $\tilde C(n,t)=\langle C(n'),t|n,t\rangle= C(n)$, which leads to a steady state distribution $ P_{st}(n)=\frac{P_{st}(0)}{\gamma^nn!}\prod_{k=0}^{n-1}C(k)$. In the non-linear case $C(n)=\frac{c}{1+\epsilon n}$, this leads to:
\begin{eqnarray}
P_{st}(n)&=&\frac{v^{\frac{1}{\epsilon}-1}}{I_{\frac{1}{\epsilon}-1}(2v)}\frac{v^{2n}}{n!\Gamma(n+\frac{1}{\epsilon})}\\
\langle n\rangle_{st}&=&v\frac{I_{\frac{1}{\epsilon}}(2v)}{I_{\frac{1}{\epsilon}-1}(2v)}\\
\sigma^2_{st}&=&\langle n\rangle_{st}-v^2\left[\left(\frac{I_{\frac{1}{\epsilon}}(2v)}{I_{\frac{1}{\epsilon}-1}(2v)}\right)^2-\frac{I_{\frac{1}{\epsilon}+1}(2v)}{I_{\frac{1}{\epsilon}-1}(2v)}\right]
\end{eqnarray}
where $v=\sqrt{\frac{c}{\gamma\epsilon}}$. It is possible to show that $\sigma^2_{st}\le\langle n\rangle_{st}$, a sub-Poissonian distribution in this case of $\tau=0$. In the next section, we will introduce an approximation that will allow us to explain that fluctuations can be amplified and become super-Poissonian when we include time-delay terms in the process.

\subsection{The time-reversal invariance assumption}

One of the difficulties for the calculation of $\tilde C_{st}(n)\equiv \lim_{t\to\infty}\langle C(n'),t-\tau|n,t)\rangle=\langle C(n'),-\tau|n\rangle_{st}$ is that it is a correlation backwards in time, whereas Eqs.(\ref{master_delay}) and (\ref{averagedelaygen}) are only valid for $t\ge 0$. The approximation we propose in this subsection is to use a time-reversal invariance assumption in the steady state, namely $\langle C(n'),-\tau|n\rangle_{st}=\langle C(n'),\tau|n\rangle_{st}$. A simple algebra shows that a sufficient condition for this time-reversal invariance to hold is that the stationary probabilities satisfy: $P_{st}(n',t|n)P_{st}(n)=P_{st}(n,t|n')P_{st}(n')$, valid for all $t\ge0$. This relation is correct in the limit $t\to 0$, as $P_{st}(n',dt|n)=w(n\to n')dt$, the rate of going from $n$ to $n'$ particles during time $dt$, and it then becomes $w(n\to n')P_{st}(n)=w(n'\to n)P_{st}(n')$, the detailed balance condition, which is valid for any one-step process, as can be derived from the master equation\cite{VK}. If the process were Markovian, the detailed balance condition would imply the time-reversal invariance for arbitrary, finite, time $t$. As the presence of delay makes the process not Markovian, the time-reversal invariance is an assumption whose validity and implications need to be checked. In Fig.~(\ref{fig:correlations}) we plot the correlations $\langle n,\tau|k\rangle_{st}$ and $\langle n,-\tau|k\rangle_{st}$ as a function of $k$, using a negative feedback loop $C(n)=\frac{c}{1+\epsilon n}$ for two different sets of parameters. In the same figure we plot the stationary probability distribution $P_{st}(k)$. As it can be seen from this figure, it is not true that these two correlations are identical for all values of $k$. However, it has to be noticed that the larger discrepancies occur for those values of $k$ which have a low probability of appearance.

\begin{figure}[h]
 \centering
 \includegraphics[width=9.5cm]{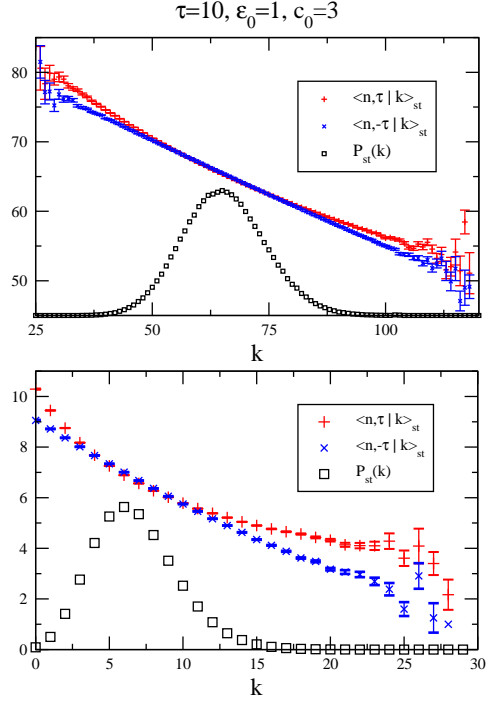}
 \caption{Conditional averages in the steady state, $\langle n,\tau|k\rangle_{st}$ (+ symbols) and $\langle n,-\tau|k\rangle_{st}$ ($\times$ symbols) coming from numerical simulations of the process with delay schematized in Eq.(\ref{eq:process}) using a creation rate $C(n)=\frac{c_0\Omega}{1+\frac{\epsilon_0}{\Omega} n}$, with $\tau=10,\epsilon_0=1, c_0=3$ and two different values of $\Omega=50$ (top) and $\Omega=5$ (bottom). In the same figures, we also plot with $\square$ symbols, the (arbitrarily rescaled) stationary probability distribution $P_{st}(k)$. Note that the discrepancy between $\langle n,\tau|k\rangle_{st}$  and $\langle n,-\tau|k\rangle_{st}$ is larger in those cases that the particular value of $k$ is less probable.
 \label{fig:correlations}}
\end{figure}

Once this time reversal invariance assumption has been adopted, to compute $\tilde C_{st}(n)=\langle C(n'),\tau|n\rangle_{st}$, one could solve the hierarchy of equations for the moments with the appropriate initial condition. This could be done, for instance, if the creation rate $C(n)$ were a linear function $C(n)=c-\epsilon n$. However, as discussed before, most interesting cases consider a negative feedback with a non-linear rate $C(n)$. In this case,  one can not, in general, close that hierarchy of equations  and one needs approximate methods to find $\tilde C_{st}(n)$, such as, for example, the Gaussian closure\cite{gaussian}. In the following, and in the spirit of van Kampen's expansion\cite{VK}, we will linearize the equations assuming  that the variable $n$ has a deterministic contribution of order $\Omega$ (a large parameter of the system, typically the system volume) and a fluctuating part of order $\Omega^{\frac{1}{2}}$ i.e. $n=\Omega \phi+\Omega^{\frac{1}{2}}\xi$. Although it is possible to deal with the most general case, we will restrict ourselves to the case where the creation rate satisfies the following scaling with system size $C(n)=\Omega \varPhi\left(\frac{n}{\Omega}\right)$, so one can expand $C(n)$ around the macroscopic component:
\begin{equation}\label{cnexpand}
 C(n)=\Omega \left(\varPhi\left(\phi\right)+\Omega^{-1/2}\varPhi'(\phi)\xi+\cdots\right),
\end{equation}
so that 
\begin{equation}
\label{eq:cvankampen}
\langle C(n'),t'|n,t\rangle=\Omega \varPhi\left(\phi(t')\right)+\Omega^{1/2}\varPhi'(\phi(t'))\langle \xi',t'|\xi,t\rangle,
\end{equation}
with  $\xi=\Omega^{-1/2}n-\Omega^{1/2}\phi$.

We replace ansatz (\ref{cnexpand}) in the evolution for the first moment (\ref{averagedelaygen}), and equate the powers of $\Omega$ to find that  the deterministic (macroscopic) and stochastic contributions to $n$ satisfy:
\begin{eqnarray}
\frac{d\phi(t)}{dt}&=&-\gamma\phi(t)+\varPhi\left(\phi(t-\tau)\right),\label{vkfluc2}\\
\frac{d\langle\xi',t'|\xi,t\rangle}{dt'}&=&-\gamma\langle\xi',t'|\xi,t\rangle+\varPhi'\left(\phi(t-\tau)\right)\langle\xi',t'-\tau|\xi,t\rangle\label{vkflucgen}
\end{eqnarray}
Equation (\ref{vkfluc2}) for the macroscopic component is, in general, a nonlinear delayed differential equation which might be difficult to solve. However, the steady state value $\phi_{st}$ is readily accessible as the solution of $\gamma\phi_{st}=\varPhi(\phi_{st})$. The stability of this fixed point is found by linearization around it. A standard analysis of the resulting linear delay differential equation, tells us that a sufficient (but not necessary) condition for stability is $|\alpha|<\gamma$, where we have defined  $\alpha\equiv -\varPhi'\left(\phi_{st}\right)$. 

Once in the steady state, we replace $\phi(t)$ by its stationary value $\phi_{st}$ and Eq. (\ref{vkflucgen}) becomes a delay linear differential equation with constant coefficients,  
and we are looking for the time-symmetric solution of this equation satisfying the initial condition $\langle\xi',t|\xi,t\rangle=\xi$. This can be written as $\langle\xi',t+\Delta|\xi,t\rangle=\xi f(\Delta)$, being $f(t)$ the symmetric solution $f(-t)=f(t)$ of the equation $\dot f(t)=-\gamma f(t)-\alpha f(t-\tau)$ and $f(0)=1$ (see Appendix 2). From Eq.(\ref{eq:cvankampen}) we get the effective creation rate $\tilde C(n)=\Omega\varPhi(\phi_{st})+\Omega^{1/2}\varPhi'(\phi_{st})\xi f(\tau)=\Omega\phi_{st}(\gamma-\varPhi'(\phi_{st})f(\tau))+\varPhi'(\phi_{st})f(\tau)n$ after replacing $\xi=\Omega^{-1/2}n-\Omega^{1/2}\phi_{st}$ and $\gamma\phi_{st}=\varPhi(\phi_{st})$. From Eq.(\ref{probgeneral}) one can obtain the steady-state probabilities $P_{st}(n)$. Their functional form depends on the sign of $\varPhi'(\phi_{st})f(\tau)$: (i) If $\varPhi'(\phi_{st})f(\tau)<0$, the distribution is a binomial distribution $P_{st}(n)={M\choose n}p^n(1-p)^{M-n}$ with $p=\frac{-\varPhi'(\phi_{st})f(\tau)}{\gamma-\varPhi'(\phi_{st})f(\tau)}$ and $M=\Omega \varPhi(\phi_{st})\left(\frac{\gamma}{-\varPhi'(\phi_{st})f(\tau)}-1\right)$ and $0\le n\le M$; (ii) if $\varPhi'(\phi_{st})f(\tau)=0$, the distribution has a Poisson form $P_{st}(n)=e^{-\chi}\frac{\chi^n}{n!}$ with $\chi=\Omega \varPhi(\phi_{st})$; (iii) finally, if $\varPhi'(\phi_{st})f(\tau)>0$, the distribution is a negative binomial, $P_{st}(n)={M+n-1\choose n}(1-q)^Mq^n$, with $q=\frac{\varPhi'(\phi_{st})f(\tau)}{\gamma}$ and $M=\Omega \varPhi(\phi_{st})\left(\frac{\gamma}{\varPhi'(\phi_{st})f(\tau)}-1\right)$. In all cases, however, they can be approximated up to terms of order $\Omega^{-1/2}$ by a Gaussian distribution. Despite the differences in the functional form, in all three cases the mean value and variance are given by:\begin{eqnarray}
 \langle n\rangle_{st}&=&\Omega\phi_{st}\label{vkaverage}\\
\sigma^2_{st}&=& \frac{\langle n\rangle_{st}}{1-\gamma^{-1}\varPhi'(\phi_{st})f(\tau)}\label{vksigma},
\end{eqnarray}
An equivalent expression for the variance taking as a starting point a linear Langevin differential equation including delay was obtained in \cite{kuchler,Frank}.

In the case of a negative feedback loop, it is $\alpha=-\varPhi'(\phi_{st})>0$. It can then be seen from the expression in the Appendix 2 that $f(\tau)$ monotonically decreases from the value $1$ at $\tau=0$ to the value $-\frac{\gamma-\lambda}{\alpha}<0$ at $\tau\rightarrow\infty$ ($\lambda=\sqrt{\gamma^2-\alpha^2}$}, see Appendix 2, recall that $|\alpha|<\gamma$ is a sufficient condition for the stability of the fixed point $\phi_{st}$). In this case  the fluctuations are sub-Poissonian if $f(\tau)>0$ (small $\tau$) and super-Poissonian if $f(\tau)<0$ (large $\tau$). The threshold between the two cases is the value $\tau_P$ at which $f(\tau_P)=0$ or $\tau_P=-\lambda^{-1}\ln\zeta$ in the notation of the Appendix 2. As explained before, the probability distribution is binomial for $\tau<\tau_P$, Poissonian for $\tau=\tau_P$ and a negative binomial for $\tau>\tau_P$.

In the case of positive feedback, $\alpha=-\varPhi'(\phi_{st})<0$, $f(\tau)$ monotonically decreases from $1$ at $\tau=0$ to $-\frac{\gamma-\lambda}{\alpha}>0$ at $\tau\rightarrow\infty$, and in this case the fluctuations are always super-Poissonian, but their magnitude is reduced as the delay is increased. The steady-state probability distribution is always a negative binomial distribution. 

We conclude that the delay can have opposite effects: in a negative feedback loop it enhances the fluctuations, whereas in a positive feedback loop it reduces them. On the other hand, it is well known that, in the non-delay scenario, a negative feedback reduces the magnitude of the fluctuations \cite{oudenardenpnas} when compared to the $n$-independent creation rate. We find it remarkable that the presence of delay can reverse the usual fluctuations-reducing effect of the negative feedback loop, and, instead, enhance the fluctuations.

The correlations in the steady state can be obtained from $K(t)=\langle n\langle n',t|n\rangle_{st}\rangle_{st}-\langle n\rangle_{st}^2$, as:
\begin{equation}
K(t)=\sigma_{st}^2f(t).\label{steadylincorr}
\end{equation}
Note that, as can be seen from the alternative definition $K(t)=\lim_{t'\to\infty}\langle n(t+t')n(t')\rangle-\langle n\rangle_{st}^2$, the correlation function is a time-symmetric function $K(-t)=K(t)$. However, and contrary to previous assumptions\cite{BVTH-2005}, this does not imply that the conditional expectation value $\langle n',t|n\rangle_{st}$ has to be a symmetric function. In fact, it is not for an arbitrary value of $n$, as shown in Fig.\ref{fig:correlations}.

We apply these results to specific functional dependences of $C(n)$. Let us first comment that in the linear case $C(n)=c-\epsilon n$, Eq.(\ref{averagedelaygen}) is already a closed equation and our treatment, not surprisingly, can be carried out without assuming the expansion (\ref{cnexpand}). However, we do not find this case very interesting as it turns out that the problem is ill-defined as the rate $C(n)$ might become negative when the number of molecules $n$ exceeds $c/\epsilon$.

A more interesting case, used in the protein transcription problem\cite{tyson}, is the rate $C(n)=\frac{c}{1+\epsilon n}$, that we write in the form $C(n)=\Omega \varPhi\left(\frac{n}{\Omega}\right)$ with $\varPhi(z)=\frac{c_0}{1+\epsilon_0 z}$ and $c_0=c/\Omega$, $\epsilon_0=\epsilon \Omega$ where $\Omega$ is a large parameter, typically proportional to the cell volume. This corresponds to a negative feedback loop. Note that the condition $|\alpha|<\gamma$ is always satisfied for such a creation rate and the steady state $\phi_{st}$ is always stable no matter how large the delay time $\tau$.

\begin{figure}[h]
 \centering
 \includegraphics[width=10cm]{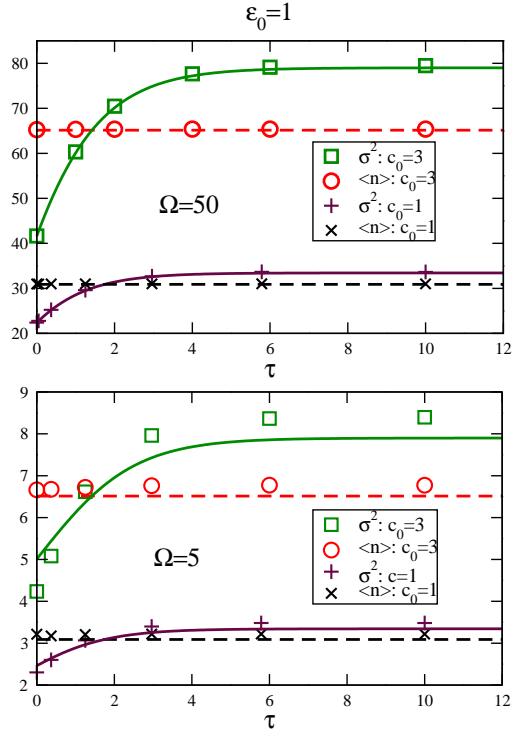}
 \caption{Steady state average $\langle n\rangle_{st}$ (dashed lines) and variance $\sigma^2_{st}$ (full lines), for process defined in  (\ref{eq:process}), as a function of the delay time $\tau$, for a creation rate $C(n)=\frac{c_0\Omega}{1+\frac{\epsilon_0}{\Omega} n}$ with $c_0=3$ (upper part of each panel) and $c_0=1$ (lower part of each panel), and two system sizes ($\Omega$) (upper and lower panel) and $\epsilon_0=1$ in both cases. In each case, we plot with symbols the results coming from numerical simulations and by lines the theoretical expressions, Eqs. (\ref{vkaverage}) and (\ref{vksigma}).
 \label{fig:phiAdd_bifur}}
\end{figure}

\begin{figure}[h]
 \centering
 \includegraphics[width=7.5cm]{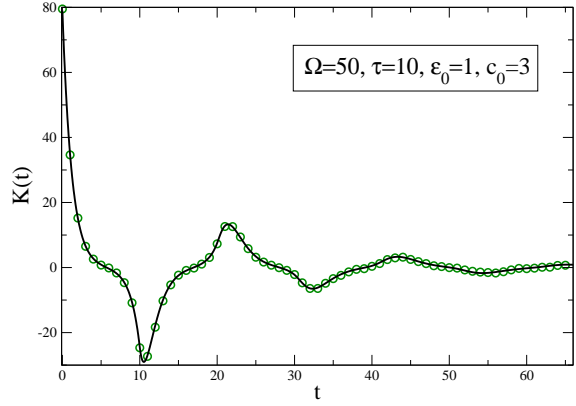}
 \caption{Correlation function in the steady state, for the delayed process (\ref{eq:process}) with creation rate $C(n)=\frac{c_0\Omega}{1+\frac{\epsilon_0}{\Omega} n}$. Simulations (circles) and theory, equation (\ref{steadylincorr}) (solid line).
 \label{fig:corrfunc}}
\end{figure}

In Fig.(\ref{fig:phiAdd_bifur}) we compare the average and variance obtained from numerical simulations with those obtained from the theoretical analysis. The agreement is, in general, very good and improves as $\Omega$ becomes large. In Fig.(\ref{fig:corrfunc}) we compare the correlation function obtained numerically with the analytical expression \ref{steadylincorr}. Its non-monotonic character due to the delay is apparent. The value of the correlation at $t=\tau$ is not negligible, compromising the validity of the independent-times approximation.

For $C(n)=\frac{c_0\Omega}{1+\epsilon_0\left(\frac{n}{\Omega}\right)^l}$ with $l>1$, (negative feedback loop with cooperativity) the equation for the macroscopic variable (\ref{vkfluc2}) has a Hopf bifurcation into a limit cycle attractor. For parameters below the Hopf bifurcation, the situation is qualitatively equal to the previous case, and the discussion applies. For parameters above the Hopf bifurcation, the system becomes oscillatory so the assumption of steady state is not valid, and the results obtained here are not directly applicable.

\section{Distributed delay}
\label{sec:3}
In general terms, it is more realistic to consider that the delay that each individual event takes to be completed is a fluctuating quantity following some probability distribution rather than taking a fixed value. This is definitely the case in genetic networks, where transcription and translation times can be broadly distributed \cite{voliotis}.
In this section we exemplify how to apply the method developed before in the case of a stochastic delayed production process including distributed delay.

We consider again the process schematized in (\ref{eq:process}), but now we consider that the delay time $\tau$ is a stochastic variable with some probability distribution $p(\tau)$. For simplicity, we consider that the delay times for all individual reactions are independent and identically distributed.
Now, the master equation for the process is:
\begin{eqnarray}
 \frac{\partial P(n,t)}{\partial t}&=&(E-1)[D(n)P(n,t)]+(E^{-1}-1)\left[\sum_{n'=0}^{\infty}\int \mathrm{d}\tau p(\tau)C(n')P(n',t-\tau;n,t)\right]\nonumber\\
&=&(E-1)[D(n)P(n,t)]+(E^{-1}-1)[\overline{C(n,t)}P(n,t)]\label{master_effdistribdelay},
\end{eqnarray}
with $\overline{C(n,t)}\equiv \int \mathrm{d}\tau p(\tau)\langle C(n'),t-\tau|n,t\rangle=\int \mathrm{d}\tau p(\tau)\tilde C(n,t;\tau)$. Hence, it is possible to follow formally the method of last section simply replacing $\tilde C$ by $\bar C$ and we skip the details of the calculation. The mean value, variance and correlation function are given by:
\begin{eqnarray}
 \langle n\rangle_{st}&=&\Omega\phi_{st},\\
\sigma^2_{st}&=& \frac{\langle n\rangle_{st}}{1-\gamma^{-1}\varPhi'(\phi_{st})\int d\tau p(\tau)f(\tau)},\label{vksigmadis}\\
K(t)&=&\sigma^2_{st}f(t),
\end{eqnarray}
being $f(t)$ the solution of the integro-differential equation
\begin{eqnarray}
\frac{df(t)}{dt}&=&-\gamma f(t)+\varPhi'\left(\phi_{st}\right)\int \mathrm{d}\tau p(\tau)f(t-\tau)\label{vkflucdelaydistrib}
\end{eqnarray}
satisfying $f(-t)=f(t)$ and $f(0)=1$.

There is no general method that can be applied to find the solution of this complicated equation. A reduction to a set of linear differential equations can be achieved if we adopt the Gamma probability distribution: $p(\tau;k)=A\tau^{k-1}e^{-\frac{k}{\overline \tau}\tau}$, depending on two parameters: $k$ and $\overline \tau$. The average value is $\overline \tau$ and the root-mean-square is $\sigma_{\tau}=\frac{\overline\tau}{\sqrt{k}}$. Increasing $k$ for fixed $\overline \tau$ decreases the fluctuations of $\tau$, and in the limit $k\rightarrow\infty$ the distribution approaches a Dirac-delta and $\tau$ becomes a deterministic variable (fixed delay, corresponding to the case analyzed in the previous section). The alternative solution method, known as the linear-chain trick\cite{smith2011}, begins by defining a family of time-dependent functions $Z_l(t)=\int \mathrm{d}\tau p(\tau;l)f(t-\tau),\,l=1,\dots,k$. After some algebra, one can prove that (\ref{vkflucdelaydistrib}) is equivalent to the system of linear ordinary differential equations:
\begin{eqnarray}
\frac{df(t)}{dt}&=&-\gamma f(t)+\varPhi'\left(\phi_{st}\right)Z_k(t),\label{xt}\\
\frac{dZ_1}{dt}&=&\frac{k}{\overline\tau}(f(t)-Z_1),\label{z1}\\
\frac{dZ_l}{dt}&=&\frac{k}{\overline\tau}(Z_{l-1}-Z_l),\hspace{1cm}l=2,\dots,k.\label{zl}
\end{eqnarray}
which, besides $f(0)=1$,  require a set of initial conditions for $Z_{l}(t=0),\,l=1,\dots,k$. These can be determined in a self-consistent manner. First, note that the symmetry condition $f(t)=f(-t)$ implies:
\begin{equation}
 Z_l(t=0)=\int \mathrm{d}\tau p(\tau;l)f(\tau),\hspace{1cm}l=1,\dots,k.\label{symmetrizl}
\end{equation}
One then solves (\ref{xt}-\ref{zl}) with arbitrary initial conditions for $Z_l(t=0)$ and imposes (\ref{symmetrizl}). This yields an algebraic system of $k$ linear equations for $Z_l(t=0)$. The solution of the linear differential equations  (\ref{xt}-\ref{zl}) and the solution of the algebraic equations (\ref{symmetrizl}) can be obtained, either analytically for small $k$, or numerically, but with a very high precision, for large $k$. Note that in order to compute the variance, Eq.(\ref{vksigmadis}), all we need to know is $\int \mathrm{d}\tau p(\tau;k)f(\tau)=Z_k(t=0)$.

\begin{figure}[h]
 \centering
\includegraphics[width=9cm]{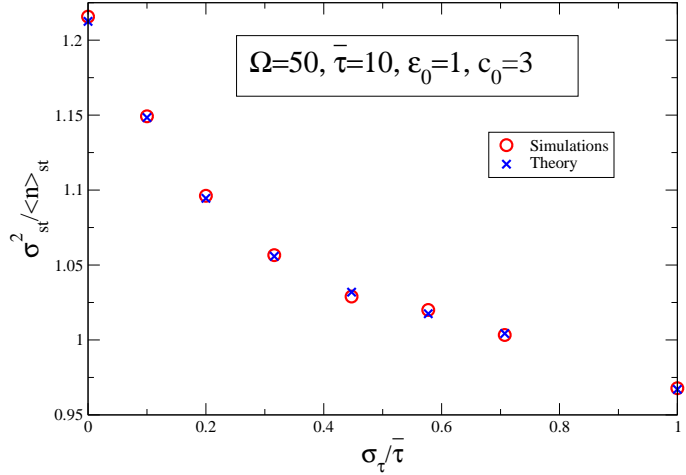}
 \caption{Variance normalized to the mean value $\sigma^2_{st}/\langle n\rangle_{st}$, for the process with distributed delay defined in (\ref{eq:process}) and a creation rate $C(n)=\frac{c_0\Omega}{1+\frac{\epsilon_0}{\Omega} n}$, for a delay distributed according to a gamma distribution $p(\tau;k)=A\tau^{k-1}e^{-\frac{k}{\overline \tau}\tau}$, as a function of the relative size of the fluctuations in the delay $\displaystyle\frac{\sigma_{\tau}}{\overline\tau}=k^{-1/2}$. Results coming from numerical simulations  ($\circ$) and from the theoretical method ($\times$) as explained in the main text. 
\label{fig:distributed}}
\end{figure}

In Fig.(\ref{fig:distributed}) we plot the ratio $\sigma_{st}^2/\langle n\rangle_{st}$ as a function of $\sigma_{\tau}$ for fixed mean delay $\overline\tau$. We see that as the delay distribution becomes wider (decreasing $k$), the fluctuations of the process decrease, so that the effect of the delay becomes less important. The results for the Gamma probability distribution are qualitatively equal to other distributions for the delay times such as uniform or Gaussian (truncated in order not to produce negative values). This ressults suggests that a natural or artificial system should have a rather precise delay if it is to make use of the effects that delay induces in the fluctuations, or it should have an irregular delay to avoid those effects.

\section{Transcription-translation model}
\label{sec:4}

So far, we have consider simple one-step birth and death processes. In the context of gene regulation, however, the protein production involves two major steps (transcription and translation) and it is well known that the combined effect of the two steps can enhance significantly protein fluctuations \cite{oudenardenpnas}. In this section we study the effect of delay in a more elaborated model for protein levels than the one considered previously, including explicitly the transcription (creation of mRNA from the DNA) and translation (creation of the protein from the mRNA) steps. The process can be schematized as follows:
\begin{eqnarray}
\emptyset {{C \atop \longrightarrow}\atop{}}Y^*{{ {}\atop \Longrightarrow}\atop{\tau_1}}Y, \hspace{0.7cm} Y{{\omega \atop \longrightarrow}\atop{}}X^* {{ {}\atop \Longrightarrow}\atop{\tau_2}}X, \hspace{0.7cm} X {{\gamma_n \atop \longrightarrow}\atop{}} \emptyset,\hspace{0.7cm} Y {{\gamma_m \atop \longrightarrow}\atop{}} \emptyset.\label{proteinreactions}
\end{eqnarray}
Now $X$ corresponds to the protein (with $n$ the current number) and $Y$ to the mRNA. We denote by $m$ the number of mRNA molecules at time $t-\tau_2$. In doing so, the translational delays $\tau_1$ and $\tau_2$ can be  absorbed in a total delay $\tau\equiv\tau_1+\tau_2$.
The master equation for the process is:
\begin{eqnarray}
 \frac{\partial P(m,n,t)}{\partial t}&=&(E_n-1)[\gamma_n nP(m,n,t)]+(E_m-1)[\gamma_m mP(m,n,t)]\label{master_delay2}\\
&+&(E_n^{-1}-1)\left[\omega mP(n,m,t)\right]+(E_m^{-1}-1)\left[\sum_{n'=0}^{\infty}C(n')P(n',t-\tau;m,n,t)\right]\nonumber
\end{eqnarray}
being $E_n$ and $E_m$ the step operators for the number of proteins, $n$,  and the number of mRNA, $m$, respectively. As before, we will allow for feedback loops by letting the creation rate $C$ to become a function on $n$. For simplicity, though, the translation rate $\omega$, as well as the degradations rates $\gamma_n$ and $\gamma_m$ will be considered constant.

\begin{figure}[h]
 \centering
 \includegraphics[width=5cm]{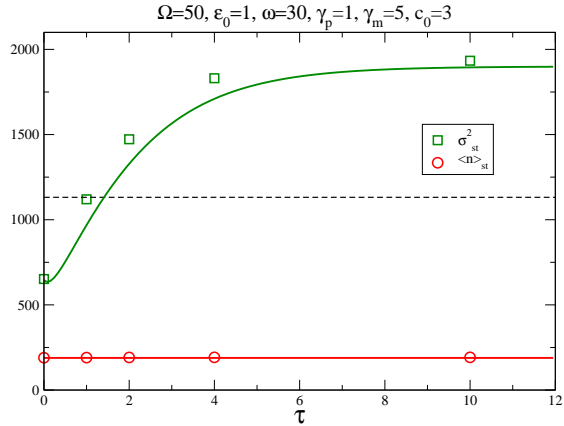}
 \caption{Stationary values for the average $\langle n\rangle_{st}$ and variance $\sigma^2_{st}$ for the protein levels as a function of the total delay, for the transcription-translation model schematized in (\ref{proteinreactions}) for a creation rate of the form $C(n)=\frac{c_0\Omega}{1+\frac{\epsilon_0}{\Omega}n}$. Values from numerical simulations (symbols) and theory (solid lines, Eq. (\ref{proteinvar})). Values of parameters in top of figure. The dashed line corresponds to the variance of a system without feedback, with the same average.
 \label{fig:transcription}}
\end{figure}

The general formal expression for the stationary solution of the master equation (\ref{master_delay2}) is not known. To proceed in this case, we will apply van Kampen's expansion, which assumes both $n$ and $m$ to be split in deterministic and stochastic contributions as $n=\Omega \phi_n+\Omega^{1/2}\xi_n$ and $m=\Omega \phi_m+\Omega^{1/2}\xi_m$. The probability density function $\Pi(\xi_n,\xi_m)$ for the stochastic variables satisfies a Fokker-Planck equation that is found by expanding the master equation in powers of $\Omega$:
\begin{eqnarray}
 \frac{\partial \Pi(\xi_m,\xi_n,t)}{\partial t}&=&\frac{\partial}{\partial\xi_m}\{\left[\gamma_m\xi_m-f'(\phi_n(t-\tau))\langle \xi_n',t-\tau|\xi_m,\xi_n,t\rangle\right]\Pi\}\nonumber\\
&+&\frac{1}{2}\left[\gamma_m\phi_m+f(\phi_n(t-\tau))\right]\frac{\partial^2}{\partial\xi_m^2}\Pi+\frac{\partial}{\partial\xi_n}\{\left[\gamma_n\xi_n-\omega\xi_m \right]\Pi\}\nonumber\\
&+&\frac{1}{2}\left[\gamma_n\phi_n+\omega\phi_m\right]\frac{\partial^2}{\partial\xi_n^2}\Pi.\label{fpgen2}
\end{eqnarray}
The deterministic contributions $\phi_n$, $\phi_m$ and the averages of the fluctuation terms obey the following system of delayed differential equations:
\begin{eqnarray}
 \frac{d\phi_m}{dt}&=&-\gamma_m\phi_m+\Phi(\phi_n(t-\tau)),\label{protmf}\\
\frac{d\phi_n}{dt}&=&-\gamma_n\phi_n+\omega\phi_m,\label{protmf2}\\
 \frac{d\langle\xi_m',t'|\xi_n,\xi_m,t\rangle}{dt'}&=&-\gamma_m\langle\xi_m',t'|\xi_n,\xi_m,t\rangle+\Phi'(\phi_n(t-\tau))\langle\xi_n',t'-\tau|\xi_n,\xi_m,t\rangle,\\
\frac{d\langle\xi_n',t'|\xi_n,\xi_m,t\rangle}{dt'}&=&-\gamma_n\langle\xi_n',t'|\xi_n,\xi_m,t\rangle+\omega\langle\xi_m',t'|\xi_n,\xi_m,t\rangle.
\end{eqnarray}
The solutions for the average of the fluctuations with appropriate initial conditions, after replacing $\phi_m(t)$ and $\phi_n(t)$ by their stationary values $\phi_{n,st}$ and $\phi_{m,st}$ coming from the fixed-point solution of Eqs.(\ref{protmf},\ref{protmf2}) can be solved under the assumption of time-reversal invariance, to obtain:
\begin{equation}
 \langle\xi_n',t|\xi_m,\xi_n\rangle_{st}=f_n(t)\xi_n+f_m(t)\xi_m
\end{equation}
(see Appendix 2 for explicit expressions of the functions $f_n(t)$ and $f_m(t)$). We replace again $\phi_m(t)$ and $\phi_n(t)$ by  $\phi_{n,st}$ and $\phi_{m,st}$  and use the time reversal approximation $ \langle\xi_n',-\tau|\xi_m,\xi_n\rangle_{st}= \langle\xi_n',\tau|\xi_m,\xi_n\rangle_{st}$ to reduce Eq.(\ref{fpgen2}) to a linear Fokker-Planck equation whose solution is well known to be a Gaussian distribution\cite{VK}. The corresponding steady state values for the average and fluctuations in protein levels are given by:
\begin{eqnarray}
 \langle n\rangle_{st}&=&\Omega\phi_{n,st}\\
\frac{ \sigma^2_{n,st}}{\langle n\rangle_{st}}&=&1+\frac{\frac{\omega}{\gamma_m}}{1+\frac{\gamma_n}{\gamma_m}+\frac{\alpha}{\gamma_m}f_m(\tau)}\frac{1-\frac{\alpha}{\gamma_n}f_n(\tau)(1+\frac{\alpha}{\gamma_m}f_m(\tau))}{1+\frac{\alpha}{\gamma_m}\left(\frac{\omega}{\gamma_n}f_n(\tau)+f_m(\tau)\right)}\label{proteinvar}
\end{eqnarray}
In the case of no delay ($\tau=0$), this expression reduces to the one obtained in \cite{oudenardenpnas}. 
In Fig.(\ref{fig:transcription}) we compare the average and variance of this transcription-translation model as a function of the delay for a creation rate of the form $C(n)=\frac{c_0\Omega}{1+\frac{\epsilon_0}{\Omega}n}$. Again, in this negative feedback loop setting, the delay significantly enhances the fluctuations, up to a level well over the value without feedback (marked in the figure by a dashed line), leaving the mean value $\langle n\rangle_{st}$ essentially unchanged. So again in this case, the delay reverts the effect of the negative feedback, from fluctuation-reducing (for low values of the delay) to fluctuation-amplifying (for large values of the delay).

\section{Discussion}
\label{sec:5}

We have studied stochastic processes with discrete variables in continuous time that include delay. 
We have shown that the combined effect of feedback and delay gives rise to nontrivial results. When a stochastic process has negative feedback, the fluctuations are decreased; if this feedback is delayed, the fluctuations can be are actually enhanced, depending on the magnitude of the delay. A positive feedback loop enhances the fluctuations, but if the feedback is delayed, this enhancement is decreased. We have also shown that this effect of the delay is less apparent if the delay itself has relative large fluctuations, so for this mechanism to work, the delay has to be controlled precisely. This may be relevant for example in gene-regulatory networks, where delay times are typically broadly distributed but several regulatory mechanisms may act to control this \cite{voliotis}. The analytical theory allows us to understand and predict this phenomenology in a general way. We have also shown that the assumption of de-correlation of times $t$ and $t-\tau$ for large delays is not justified a priori, since the correlation function is typically non-monotonically decreasing, with peaks at multiples of the delay. Finally, we have pointed out that systems with delay are not, in general, statistically invariant under time reversal over the steady state, even if they fulfill the detailed balance condition.

\section*{Appendix 1: numerical simulations}
To perform numerical realizations of the process, we use the following modification of the Gillespie algorithm \cite{gillespie,cai}:

1: Initialize the state of the system, setting, e.g. $n=0$. 

2: Compute the reaction rates $C(n)$ and $\gamma n$. Obtain a number $\Delta t$ exponentially distributed with average $1/(C(n)+\gamma n)$.

3: If $t+\Delta t$ is larger than the time of the next scheduled delayed reaction, go to step 4. Otherwise, update time from $t$ to $t+\Delta t$ and obtain which kind of process (creation or degradation) will take place. To do so, generate a uniform random number between $0$ and $1$. If this number is smaller than $\gamma n/(C(n)+\gamma n)$, set $n\to n-1$; otherwise add an entry in the list of scheduled creation processes to happen at time $t+\tau$. Go to step 2.

4: Update the time to that of the next scheduled reaction. Set $n\to n+1$. Go to step 2.

This procedure is statistically exact, as the original Gillespie algorithm in the case of non-delayed reactions.

In the case with delay, the time until the next reaction is exponentially distributed, with average $C(n)+\gamma n$, only if the state of the system doesn't change during this interval (due to a scheduled delayed reaction). This happens with probability $1-e^{-(C(n)+\gamma n)t_\tau}$ (with $t+t_\tau$ the time of the next scheduled delayed reaction). The algorithm fulfills this, since the probability that step 3 is completed is precisely $1-e^{-(C(n)+\gamma n)t_\tau}$. Once a reaction has taken place (delayed or not) the time for the next reaction is again exponentially distributed as long as no delayed reaction takes place, and the procedure can be iterated.   

\section*{Appendix 2: solution of the delay-linear equations}
We consider the following linear delayed differential equation: 
\begin{equation}
 \frac{df(t)}{dt}=-\alpha f(t-\tau)-\gamma f(t)\label{ldde}.
\end{equation}
We are looking for a symmetric solution $f(-t)=f(t)$. We summarize here for completeness the treatment of reference \cite{BVTH-2005}. We make the ansatz $f(t)=ae^{\lambda |t|}+be^{-\lambda  |t|}$, valid only for $-\tau\le t\le\tau$. Inserting in (\ref{ldde}), equating the coefficients of $e^{\lambda t}$ and $e^{-\lambda t}$, and imposing $f(0)=1$, we obtain $\lambda,a,b$. Once we know $f(t)$ for $|t|\le\tau$, we can obtain $f(t)$ for $|t|>\tau$ iteratively integrating (\ref{ldde}). The solution for $t\ge0$ is: 
\begin{eqnarray}
\lambda&\equiv&\sqrt{\gamma^2-\alpha^2}, \hspace{1cm}\zeta\equiv\frac{\gamma-\lambda}{\alpha},\nonumber\\
f(t)&\equiv&
\begin{cases}\frac{e^{-\lambda t}-\zeta e^{\lambda (t-\tau)}}{1-\zeta e^{-\lambda\tau}},&\text{if }0\le t\le\tau\\ \\
e^{-\gamma (t-k\tau)}f(k\tau)-\alpha\int_{k\tau}^tdt'\,f(t'-\tau)e^{\gamma (t'-t)},&\text{if }k\tau \le t\le(k+1)\tau, k=1,2,\cdots
\end{cases}
\end{eqnarray}
Note that $\displaystyle f(\tau)=\frac{e^{-\lambda \tau}-\zeta}{1-\zeta e^{-\lambda\tau}}$. Using the symbolic manipulation program Mathematica\cite{mathematica} to perform the integrals of the iterative process, we have been able to find explicit expressions for $f(t)$ up to $|t|\le 10 \tau$.

We apply a similar approach to  the case of two coupled linear delayed differential equations:
\begin{eqnarray}
 \frac{dx_m(t)}{dt}&=&-\gamma_mx_m(t)-\alpha x_m(t-\tau),\\
\frac{dx_n(t)}{dt}&=&-\gamma_nx_n(t)+wx_m(t).
\end{eqnarray}
Due to the linearity, the solution has the form:
\begin{equation}
 x_n(t)=x_n(0)f_n(t)+x_m(0)f_m(t),
\end{equation}
with $f_n(0)=1$, $f_m(0)=0$. To find this solution, we use the ansatz $x_n(t)=a_1e^{\lambda_+ |t|}+b_1e^{-\lambda_+ |t|}+a_2e^{\lambda_- |t|}+b_2e^{\lambda_- |t|}$, $x_m(t)=c_1e^{\lambda_+ |t|}+c_2e^{-\lambda_+ |t|}+d_1e^{\lambda_- |t|}+d_2e^{\lambda_- |t|}$, for $-\tau\le t\le\tau$. Equating the coefficients of the exponentials and imposing the initial condition we obtain the expression valid in $0\le t\le \tau$:
\begin{eqnarray}
 f_n(t)&=&\left[\gamma_n\frac{1-b_-(t)}{b(t)}+\lambda_-\frac{1+b_-(t)}{b(t)}\right]\left(e^{\lambda_+t}-b_+(t)e^{-\lambda_+t}\right)\\
 & &-\left[\gamma_n\frac{1-b_+(t)}{b(t)}+\lambda_+\frac{1+b_+(t)}{b(t)}\right]\left(e^{\lambda_-t}-b_-(t)e^{-\lambda_-t}\right),\\
f_m(t)&=&\omega\frac{1-b_+(t)}{b(t)}\left[e^{\lambda_-t}-b_-(t)e^{-\lambda_-t}\right]-\omega\frac{1-b_-(t)}{b(t)}\left[e^{\lambda_+t}-b_+(t)e^{-\lambda_+t}\right],\\
\lambda_{\pm}&=&\sqrt{\frac{\gamma_m^2+\gamma_n^2}{2}\pm\frac{1}{2}\sqrt{(\gamma_m^2-\gamma_n^2)^2+4\omega^2\alpha^2}},\\
b_{\pm}(t)&=&\frac{\lambda_{\pm}^2+(\gamma_m+\gamma_n)\lambda_{\pm}+\gamma_n\gamma_m}{\omega\alpha}e^{\lambda_{\pm}t},\\
b(t)&=&\lambda_{-}(1+b_-(t))(1-b_+(t))-\lambda_+(1-b_-(t))(1+b_+(t)).
\end{eqnarray}

{\textbf{Acknowledgments:}}
We thank J. Garcia-Ojalvo and J. Buceta for useful discussions. We acknowledge financial support by the MICINN (Spain) and FEDER (EU) through project FIS2007-60327. L.F.L. is supported by the JAEPredoc program of CSIC.


\end{document}